\documentclass[11pt]{article}

\usepackage[utf8]{inputenc}
\usepackage[T1]{fontenc}
\usepackage{lmodern}
\usepackage{amsmath,amssymb}
\usepackage{graphicx}
\usepackage{hyperref}
\usepackage{geometry}
\usepackage{setspace}
\usepackage{indentfirst}
\usepackage{booktabs}

\usepackage{makecell}
\usepackage{xcolor}

\usepackage{graphicx}
\usepackage{tikz}
\usetikzlibrary{calc} % for [xshift=...,yshift=...] syntax with anchors

\begin{document}

\title{\textbf{Targeted cooling of urban cycling networks for heat-resilient mobility}}

\author{
Agustin Cabrera$^{1}$, David Ziegler$^{1,2}$, and Markus Schläpfer$^{1,\ast}$\\[0.25em]\\
\small $^{1}$Department of Civil Engineering and Engineering Mechanics, Columbia University, New York, NY 10027, USA\\
\small $^{2}$School of Engineering and Design, Technical University of Munich, 80333 Munich, Germany\\
\small $\ast$Email: \texttt{m.schlaepfer@columbia.edu}
}

\maketitle

\textbf{Cities are increasingly challenged by extreme heat events, which pose serious risks to public health and urban livability. Micromobility users, whose numbers have increased rapidly in recent years, are particularly vulnerable to outdoor heat exposure. Yet, their exposure patterns and the effectiveness of mitigation measures remain poorly understood. Here, we couple a high-resolution urban microclimate model (WRF--BEP--SOLWEIG) with 4.76 million Citi Bike trips in New York City to quantify cyclists’ thermal exposure during the June 2024 heatwave and to evaluate targeted cooling strategies. Results show that a small fraction of the street network concentrates the majority of rider heat exposure, and that localized interventions along these segments yield the greatest benefits. Targeted tree planting along just 1.5\% of the city’s street network reduces total heat-exposed kilometers ridden by 19\%, equivalent to a thermal stress reduction of about 4\,$^{\circ}$C, with its impact maximized during midday hours. In contrast, randomized citywide tree planting produces diffuse, resource-intensive cooling, highlighting the superior efficiency of spatially prioritized interventions. Baseline results further indicate that daytime heat stress is higher in lower-income neighborhoods, adding an important social dimension of urban heat exposure. Together, these findings provide a quantitative basis for designing heat-resilient and equitable cycling networks in a warming climate.}\newpage
%\smallskip
%\noindent\textbf{Key words:} Urban heat exposure, Micromobility, Thermal comfort, UTCI, Urban climate modeling, Cycling infrastructure

% ----------------------------------------------------------
% 1. Introduction
% ----------------------------------------------------------
% \section{Introduction}

Extreme heat is becoming increasingly frequent and intense, and in cities it is further amplified by the urban heat island effect~\cite{ref1}, posing a growing threat to public health. In the United States and Europe, heat stress causes more premature deaths than any other climate%
-related hazard~\cite{ref2}. At the same time, many cities are promoting micromobility options, such as bicycles and e-bikes, to enhance accessibility and encourage sustainable transportation~\cite{ref3}. In New York City, for example, the ``Citi Bike'' bicycle-sharing system recorded over 44 million trips in 2024, setting a new annual record~\cite{ref4}. This growing reliance on outdoor, human-powered transport introduces a new dimension to urban heat exposure that remains largely unaddressed in existing adaptation strategies. It underscores the need to protect micromobility users from extreme heat through mitigation measures, such as providing shade, adding vegetation, or installing ``cool pavements''~\cite{ref5,ref6}.

Trees and other forms of urban vegetation have shown strong potential to mitigate both the urban heat island effect and localized heat stress by reducing the Universal Thermal Climate Index (UTCI), which describes how the human body experiences atmospheric conditions, specifically air temperature, humidity, wind, and radiation~\cite{ref7}. Prior studies demonstrate that shaded and vegetated routes substantially improve cyclists’ thermal environments. Lanza et al.\ (2025)~\cite{ref8} found that cycling on vegetated trails lowered air temperature and light intensity relative to roadways, while Garber et al.\ (2025)~\cite{ref9} reported that green and blue spaces in Denver, USA, reduced the negative impact of heat on cycling activity, supporting the value of tree-lined ``cool corridors.'' Rodriguez et al.\ (2025)~\cite{ref10} further quantified daytime UTCI reductions of 6--10\,$^{\circ}$C under shaded street canopies. A global meta-analysis by Li et al.\ (2024)%
~\cite{ref11} confirmed that the cooling efficacy of trees is greatest in open urban morphologies and temperate climates, reinforcing their universal importance for urban heat mitigation. Collectively, these studies demonstrate that strategically planting trees along urban corridors can reduce cyclists’ heat exposure while enhancing overall urban thermal resilience.

However, a limited quantitative understanding remains, not only of how micromobility users experience and can be protected from urban heat, but also of how many trees are required to meaningfully improve cyclists’ thermal comfort across an entire city during extreme heat events. Existing city-scale studies on heat exposure typically rely on static representations of human activity, such as home locations~\cite{ref12,ref13} or coarse-grained place visitation patterns~\cite{ref14}, which overlook detailed movement paths and do not distinguish between individuals’ outdoor and indoor activities. Other recent studies use pedestrian flow modeling~\cite{ref15,ref16} or GPS-based trajectories of a small number of local food delivery riders~\cite{ref17}, but these approaches are typically restricted to single neighborhoods and therefore cannot represent micromobility users who traverse multiple urban areas.

From an urban microclimate perspective, a further limitation of existing studies arises from the environmental data used. Many analyses depend on satellite-derived land surface temperature~\cite{ref12,ref13,ref18}, which cannot be directly translated into street-level air temperatures%
~\cite{ref19}. Others rely solely on solar radiation~\cite{ref15,ref16,ref17}, which limits their ability to assess conditions with little or no sunlight, such as during nighttime hours. Together, these constraints underscore the need for integrated modeling frameworks that can capture both human mobility patterns and the fine-scale spatial and temporal variability of urban thermal environments.

In this paper, we address these gaps by coupling a state-of-the-art, high-resolution urban thermal comfort model with the trajectories of shared bike users recorded during the June 2024 heatwave in New York City (NYC). Our coupled WRF--BEP--SOLWEIG framework integrates mesoscale climate simulations with microclimate processes, enabling the fine-grained analysis of diverse factors beyond solar radiation, such as oceanic influences and nighttime cooling patterns. We focus on a set of targeted interventions designed to identify and cool the routes where cyclists experience the highest thermal stress. To that end, millions of Citi Bike trips were mapped and overlaid with 10\,m--resolution UTCI maps generated by the WRF--BEP--SOLWEIG framework. Street segments were subsequently prioritized using two criteria: (1) a UTCI 
threshold of 32\,$^{\circ}$C, corresponding to strong heat stress~\cite{ref7}, and (2) rider frequency, indicating the most heavily used routes. We then simulated both the idealized effects of temperature reductions and the realistic effects of tree planting along these critical road segments. Results show that targeted interventions applied to a small subset of the city’s street network can yield substantial reductions in rider heat exposure, far outperforming non-targeted tree planting. Finally, by integrating socio-economic data, we show that riders traveling through lower-income neighborhoods experience higher daytime UTCI values, extending previous work on heat exposure inequality to the domain of micromobility.

% ----------------------------------------------------------
% 2. Methodology
% ----------------------------------------------------------
\section{Methodology}

\subsection{Study area}

The study area is New York City (NYC)
%, defined by the polygon bounded by 
%$40.615^{\circ}\mathrm{N}$, $74.388^{\circ}\mathrm{W}$; $40.480^{\circ}\mathrm{N}$, $73.810^{\circ}\mathrm{W}$; 
%$40.919^{\circ}\mathrm{N}$, $73.629^{\circ}\mathrm{W}$; and $41.055^{\circ}\mathrm{N}$, $74.211^{\circ}\mathrm{W}$ 
(see Supplementary Fig.~1), home to North America’s largest bicycle network, 
encompassing 1,550 lane miles, with 99\% of residents living within one mile of a designated bicycle route~\cite{ref20}. The city is also highly vulnerable to extreme heat, with an average of more than 500 New Yorkers dying prematurely each summer due to hot weather~\cite{ref21}.

\subsection{WRF--BEP--SOLWEIG framework}

\subsubsection{WRF--BEP model description}

To model local climate conditions, we applied a recently developed approach that couples a mesoscale atmospheric model with an urban microclimate model~\cite{ref22}. Specifically, this state-of-the-art framework integrates the Weather Research and Forecasting model with the Urban Canopy Model (WRF--UCM) and the Solar and Longwave Environmental Irradiance Geometry 
model (SOLWEIG).

The WRF model (version 4.6.0), integrated with the Building Effect Parameterization (BEP) scheme, was used to generate meteorological inputs for the study area. The BEP scheme captures the interactions between buildings and the urban boundary layer by representing the three%
-dimensional structure of the city. It accounts for the influence of vertical and horizontal surfaces on airflow, temperature, and turbulence, while also estimating heat emissions from the urban canopy through the combined effects of drag forces, diffusion, and radiative processes~\cite{ref23}.

% \subsubsection{Model configuration}

For this study, the innermost WRF--BEP domain was refined using a hybrid 100\,m global land%
-cover dataset that integrates Local Climate Zones (LCZ) from the Copernicus Global Land Service (CGLC) and the MODIS IGBP product~\cite{ref24}. This dataset provides detailed urban classifications, improving the model’s representation of NYC’s heterogeneous built environment. The WRF--BEP scheme was configured to incorporate key parameters that describe urban morphology, including building height distributions, urban fraction, albedo, and surface emissivity, thereby capturing the city’s three-dimensional structure. Building height distributions were estimated from LCZ data produced by the LCZ Generator~\cite{ref25} and calibrated with building footprint information from NYC OpenData \url{(https://data.cityofnewyork.us/City-Government/BUILDING/5zhs-2jue/about_data)}. These updates enhance the model’s ability to simulate fine-scale variations in surface--atmosphere interactions across diverse urban forms. Full configuration details and parameter sources are provided in the Supplementary Information.

%\subsubsection{Simulation period and initial boundary conditions}

The simulation employed three two-way nested domains (Supplementary Fig.~1) with horizontal resolutions of 12.5\,km, 2.5\,km, and 0.5\,km, respectively. Each domain consisted of a $101 \times 101$ grid matrix, where each cell size matched the domain’s horizontal resolution. The simulation covers a two-day heatwave period, spanning from June 19, 2024, at 8:00\,p.m.\ EDT to June 21, 2024, at 8:00\,p.m.\ EDT. The first twenty-four hours of the simulation were treated as the model spin-up period, allowing for atmospheric stabilization, and were excluded from the analysis (details in Supplementary Table~S1). Initial and boundary conditions were taken from the six-hourly National Centers for Environmental Prediction (NCEP) Global Forecast System (GFS) reanalysis data on a $1^{\circ} \times 1^{\circ}$ grid~\cite{ref26}.

\subsubsection{SOLWEIG model}

Following the mesoscale simulation with the WRF--BEP model at 500\,m resolution, the resulting meteorological outputs were used as inputs for the Solar and Longwave Environmental Irradiance Geometry (SOLWEIG) microscale model~\cite{ref27} to assess detailed urban heat conditions. SOLWEIG was applied at 10\,m spatial resolution within each WRF grid cell to compute the mean radiant temperature (MRT), which integrates short- and long-wave radiation fluxes from the sky, ground, buildings, and vegetation, expressed as an equivalent temperature~\cite{ref22,ref15}.

Input data for SOLWEIG included hourly WRF-derived meteorological variables and detailed urban morphological datasets, all resampled to a uniform 10\,m grid to capture fine-scale spatial variability across the city. A simplified canopy layer based on the land-cover ``Tree'' category was added, distributing virtual trees along streets to capture their local cooling influence. Each tree was modeled with consistent geometry to isolate the influence of added shade, and the underlying surfaces were set to grass cover following standard model assumptions~\cite{ref22}.

Note that the modeled canopy layer extends over the East River bridge corridors. While extensive tree installation on bridges is not physically feasible due to space constraints, these modeled effects should be interpreted as proxies for functionally equivalent cooling measures (e.g., shading structures, reflective surfacing, or misting systems). The hourly 10\,m fields produced by SOLWEIG form the basis for UTCI computation (Section~\ref{sec:utci}) and subsequent exposure analysis (Sections~\ref{sec:preprocess}--\ref{sec:validation}).

\subsection{UTCI calculation}
\label{sec:utci}

The Universal Thermal Climate Index (UTCI)~\cite{ref28} was computed for each 10\,m cell to evaluate outdoor thermal comfort at high spatial resolution. UTCI values were derived using the sixth-order polynomial approximation~\cite{ref22} from air temperature, relative humidity, and wind speed (from WRF--BEP) together with mean radiant temperature (MRT) data (from SOLWEIG). This combined approach enables a detailed assessment of microclimatic thermal stress across the urban environment.

\subsection{Street network preprocessing and trip routing}
\label{sec:preprocess}

The street network data were acquired from OpenStreetMap (OSM) \url{(https://wiki.openstreetmap.org/wiki/Downloading_data; https://extract.bbbike.org/)}
using the OSMnx Python package. Routing computations were performed through Valhalla \url{(https://github.com/valhalla/valhalla)}, 
and the resulting routes were simplified with the NEATNet~\cite{ref29,ref30} framework to ensure computational efficiency while preserving network realism. The resulting network totals 10{,}090\,km of rideable segments.

Citi Bike trip records for June 2024 were obtained from the official monthly operating report%
~\cite{ref31}, comprising approximately 4.76 million trips across New York City. Rather than computing unique routes for each individual trip, all trips were matched to pre-computed shortest paths between Citi Bike stations on a simplified OSM-based street network of New York City. This approach ensured consistency across trips while maintaining realistic routing patterns. Each street segment was then spatially overlaid on the hourly 10\,m UTCI raster maps, and the mean UTCI value was calculated based on all cells intersecting the segment. These averaged segment values established the baseline exposure used in the scenario analysis (see Figure~1).

\subsection{Design of scenarios}

To design interventions, hot street segments were identified based on thermal thresholds. For each hour of the day, all segments with mean UTCI values above 32\,$^{\circ}$C were flagged as heat-exposed. Overlaying trip routes onto these segments produced cumulative ranked lists of the 150, 400, and 1{,}000 most traversed heat-exposed street segments (see Figure~2). Two types of cooling interventions were then modeled to evaluate potential mitigation strategies under both idealized and realistic conditions. The targeted cooling scenarios are defined as follows:
\begin{itemize}
    \item \textbf{Hypothetical cooling.} UTCI values were hypothetically reduced by 3\,$^{\circ}$C, 5\,$^{\circ}$C, and 10\,$^{\circ}$C for the top 150, 450, and 1{,}000 most heat-exposed street segments, resulting in a total of nine scenarios. The full analysis pipeline was re-run for every case, generating updated exposure metrics and allowing direct comparison across interventions.
    \item \textbf{Tree planting.} To simulate a land-cover--based mitigation strategy, all cells within 20\,m buffers of the top 1{,}000 most traversed hot street segments (UTCI $\geq$ 32\,$^{\circ}$C) were converted into tree cells (see Figure~3). The SOLWEIG model was then re-run using this modified land cover to produce new UTCI maps that reflect the cooling effect of added vegetation. These updated maps were passed through the exposure pipeline to quantify the cooling potential and spatial redistribution of thermal comfort resulting from tree canopy expansion.
\end{itemize}

\subsection{Heat stress quantification}

We assume that cyclists take the shortest path between their origin $i$ and destination $j$,
\begin{equation}
\Pi^{\ast}_{i \rightarrow j} = 
\operatorname*{arg\,min} \bigl[ L_{i \rightarrow j}(\Pi_{i \rightarrow j}) \bigr],
\end{equation}
where
\begin{equation}
L_{i \rightarrow j}(\Pi_{i \rightarrow j}) = \sum_{e \in \Pi_{i \rightarrow j}} \ell_e
\end{equation}
represents the total length across all road segments $e$ in the path $\Pi_{i \rightarrow j}$~\cite{ref32}. This assumption is reasonable, as shown in the literature on cycling route choice behavior~\cite{ref33}.

To calculate the experienced average heat stress $H_{i \rightarrow j}$ (in $^{\circ}$C) of an individual traveling from $i$ to $j$, we assume a constant average cycling speed $v$. $H_{i \rightarrow j}$ is assumed proportional to the exposure time to the thermal comfort index $\mathrm{UTCI}(\mathbf{x}, t)$ along the travel trajectory and calculated for each location $\mathbf{x}$ and time $t$ as
\begin{equation}
H_{i \rightarrow j} =
\frac{\displaystyle \int_{\Pi^{\ast}_{i \rightarrow j}} \mathrm{UTCI}(\mathbf{x}, t)\, \mathrm{d}s}
{\displaystyle L_{i \rightarrow j}(\Pi_{i \rightarrow j})}.
\label{eq:heat_stress}
\end{equation}

\subsection{Routing and model validation}
\label{sec:validation}

To assess the reliability of the modeled exposure counts, simulated trip totals were validated against independent counter data from the NYC Department of Transportation (DOT) \url{(https://newyorkcitydot.eco-counter.com/?year=2024&month=6&siteId=100009427)},
which records bicycle volumes across the city’s major bridges. The comparison focused on the four primary East River bridges, where consistent ridership monitoring is available. According to DOT \url{(https://www.nyc.gov/html/dot/html/bicyclists/bikestats.shtml\#highlights)},
New York City averages approximately 620{,}000 cycling trips per day, with Citi Bike trips accounting for around 159{,}000 daily rides, roughly 25\% of the total~\cite{ref31}. When compared with observed bridge counts, our modeled results captured 16--30\% of total daily bicycle volumes. This range is consistent with expectations, since the model represents only Citi Bike trips and therefore should capture a proportion close to their citywide share. Moreover, the analysis was restricted to heat-exposed street segments (UTCI $\geq 32^{\circ}$C), which further narrows the sample. Taken together, this agreement indicates that the modeling framework accurately reflects both the spatial and volumetric distribution of bicycle activity across the city.

Beyond route-level validation, the reliability of the WRF--BEP--SOLWEIG framework is well established in prior studies across multiple cities and climates, including Singapore (Mughal et al., 2019; Mughal, 2020)~\cite{ref34,ref35}, Metro Manila (Bilang et al., 2022)~\cite{ref36}, and Barcelona (Ribeiro et al., 2021; Segura et al., 2021)~\cite{ref37,ref38}. These evaluations consistently demonstrate the model’s ability to reproduce urban heat island dynamics, lending confidence to its application for NYC’s complex urban environment.

% =========================================================
% 3. RESULTS
% =========================================================
\section{Results}

\subsection{Targeted interventions: identifying and cooling high-exposure segments}

We begin by presenting the results of the targeted cooling interventions, which constitute the central and most policy-relevant findings of this study. Rather than first examining citywide averages or randomized scenarios, we directly assess where cyclists experience the most severe heat exposure and how localized mitigation can most efficiently reduce that burden. This section identifies the primary network hotspots, quantifies their contribution to overall exposure, and evaluates the potential of targeted interventions to deliver maximum cooling impact with minimal resource use.

\subsubsection{Network hotspots and baseline concentration}

The preceding modeling framework allows us to pinpoint where cyclists experience the greatest thermal burden across New York City. Using the baseline UTCI values (without interventions), we identify a small set of high-exposure segments that disproportionately contribute to total rider heat stress. Across the full dataset, Citi Bike trips accumulated 8.33 million heat-exposed kilometers (trips on segments with mean UTCI $\geq 32^{\circ}\mathrm{C}$). The top 150, 400, and 1{,}000 most heat-exposed segments comprise only 34.48\,km (0.34\%), 61.09\,km (0.61\%), and 150.80\,km (1.49\%) of the analyzed total network, yet they account for 27.4\%, 35.4\%, and 51.4\% of all heat-exposed kilometers ridden, respectively (see Table~\ref{tab:scenarios}). 
These hotspots cluster along the Hudson River Greenway, Park Avenue, the East River bridges, and sections of Broadway and 8th Avenue (see Figure~2), indicating that a very limited share of the street network carries most of the exposure burden. This spatial concentration provides the basis for evaluating targeted cooling strategies in the following sections.

\begin{table}[t]
\centering
\caption{Design of theoretical cooling scenarios targeting the most heat-exposed and frequently used segments (UTCI $\geq 32^{\circ}\mathrm{C}$). Street segments were ranked by ridership frequency, and cumulative thresholds (150, 400, and 1{,}000 segments) were used to define scenario extents.}
\vspace{0.6em}   % adjust amount as needed
\label{tab:scenarios}

\begin{tabular}{lccc}
\hline
Scenario & Length (km) & \% of street network & 
\makecell{Heat-exposed km ridden \\ (\% of all heat-exposed km)} \\
\hline
Top 150 segments     & 34.48  & 0.34 & 27.40 \\
Top 400 segments     & 61.09  & 0.61 & 35.40 \\
Top 1{,}000 segments & 150.80 & 1.49 & 51.36 \\
\hline
\end{tabular}
\end{table}

\subsubsection{Effects of cooling interventions on heat-exposed kilometers ridden}

We evaluated ten segment-focused cooling scenarios combining hypothetical temperature reductions of 3$^{\circ}$C, 5$^{\circ}$C, and 10$^{\circ}$C applied to the top 150, 400, and 1{,}000 most heat-exposed street segments. Cooling the top 1{,}000 segments by 10$^{\circ}$C reduces total heat-exposed kilometers by $\sim$51\%. There are clear diminishing marginal returns: for example, a 10$^{\circ}$C cooling of the top 150 segments is more effective than a 5$^{\circ}$C cooling of the top 1{,}000 segments. In the realistic tree-%
planting scenario, adding canopy within the 1{,}000 hottest segments lowered heat-exposed kilometers by 19\%, implying an approximate $\sim$4$^{\circ}$C theoretical cooling equivalent along these segments.

\subsection{Rider outcomes: heat danger-zone share and thermal load}

Next, we translate these network-level reductions into rider outcomes by examining two rider-centric metrics: the heat danger-zone share of trips ($\geq 32^{\circ}$C and $\geq 38^{\circ}$C) and the average heat stress (see Equation~\ref{eq:heat_stress}), to evaluate how interventions improve conditions during peak riding hours (see Figure~4).

The UTCI thresholds of 32$^{\circ}$C and 38$^{\circ}$C were determined from Bröde et al.\ (2012)~\cite{ref39}. According to their criteria, around 32$^{\circ}$C, conditions correspond to strong heat stress, characterized by rising core and skin temperatures, active sweating, and increased cardiovascular and evaporative cooling demand as the body intensifies its thermoregulatory effort. Near 38$^{\circ}$C, very strong heat stress occurs, accompanied by further elevation of core temperature and skin wettedness with diminishing heat-loss efficiency, indicating that the 
body’s cooling mechanisms are approaching their physiological limits and the risk of heat exhaustion becomes significant.

Under baseline conditions, Citi Bike riders are routinely exposed to such stress levels: over 75\% of trips occur above 32$^{\circ}$C, and about one-third exceed 38$^{\circ}$C during daytime hours (08:00--%
19:00). The average heat stress reaches 35$^{\circ}$C during the day, approaching the ``strong heat stress'' category.

Cooling interventions substantially reduce these exposures. The 10$^{\circ}$C cooling applied to the top 1{,}000 segments produces the largest improvement, cutting the share of trips above 32$^{\circ}$C by nearly 20 percentage points and reducing the mean daytime UTCI from 35.1$^{\circ}$C to 32.2$^{\circ}$C (see 
Figure~S8 in the Supplementary Information), effectively below the danger threshold. Tree planting yields comparable benefits, achieving reductions equivalent to a theoretical 3--5$^{\circ}$C cooling along the most heat-exposed routes.

Average thermal load follows a similar trend. Large-scale or high-intensity cooling not only decreases the share of dangerous trips but also lowers the cumulative rider heat burden across the city. The 10$^{\circ}$C cooling of the top 1{,}000 segments yields a drop of $\sim$3$^{\circ}$C in the network-wide 
average heat stress, while the tree-planting scenario achieves a $\sim$1.5$^{\circ}$C reduction.

Together, these findings demonstrate that targeted interventions along priority segments can appreciably lower both the intensity and duration of heat exposure, reducing the share of unsafe trips and pulling average rider conditions closer to thermal comfort levels during the city’s hottest periods.

\subsection{Diurnal variation in rider thermal load}

Hourly average heat stresses reveal a clear daily pattern in rider heat exposure (see Figure~5). Under baseline conditions, average rider UTCI decreases through the early morning (23~$\rightarrow$ 21$^{\circ}$C between 00:00--06:00) and rises sharply to a midday peak of 38--39$^{\circ}$C (11:00--15:00) before declining again in the evening.

The tree-planting scenario modifies this cycle by introducing mild nocturnal warming (+0.3--0.4$^{\circ}$C) from canopy heat storage, but also strong daytime cooling of 1.5--2$^{\circ}$C between 09:00 and 16:00. Compared to the artificial cooling tests on the top 1{,}000 segments, this corresponds roughly to a 5$^{\circ}$C reduction during midday and about 3$^{\circ}$C during early morning and evening. The resulting hourly variation in cooling intensity is illustrated in Figure~5. Artificial cooling, in contrast, produces steady reductions across all hours without changing the diurnal shape. For instance, cooling the top 1{,}000 segments by 10$^{\circ}$C lowers the midday UTCI to $\sim$35$^{\circ}$C,compared with 38--39$^{\circ}$C in the baseline.

Overall, these results emphasize two points: (1) rider exposure peaks sharply at midday when both heat and trip activity are high, and (2) tree planting creates time-dependent effects: cooling riders most when it matters, during the day, while causing slight nocturnal warming, offering a realistic representation of nature-based thermal adaptation in dense urban settings.

\subsection{Citywide random tree-planting scenarios}

While the targeted segment interventions achieve substantial and concentrated reductions in rider heat exposure, it is also instructive to compare them with a non-targeted greening strategy. To contextualize their relative efficiency, we modeled a set of citywide tree-planting scenarios in which canopy cover was uniformly increased across New York City without spatial prioritization. 
These experiments illustrate the limits of diffuse, resource-intensive cooling measures and help quantify the added value of the targeted approach presented above. For this analysis, we used a subset of the Citi Bike dataset covering approximately $5.7\times 10^{5}$ trips during the peak of the June 2024 heatwave (17--23 June).

In line with previous studies~\cite{ref40}, uniformly increasing tree cover across the city lowers daytime UTCI values but with diminishing overall benefit relative to the number of trees added (see Figure~6). Adding 50\% more trees (1.5$\times$ scenario) decreases the midday UTCI peak from roughly 39$^{\circ}$C to 38$^{\circ}$C, while tripling the total canopy (3$\times$ scenario) lowers it further to $\sim$35.5$^{\circ}$C. The cooling effect is slightly stronger in low-income neighborhoods, narrowing socioeconomic 
differences in daytime thermal stress and causing the diurnal UTCI curves to converge toward a common profile. Moreover, additional trees slightly elevate nighttime UTCI due to reduced wind speed and limited long-wave heat release, a sheltering effect also reported in prior work%
~\cite{ref40}. 

However, the cooling gains remain modest when considering the large increase in canopy required. These results suggest that randomly distributed tree planting provides broad but inefficient cooling, highlighting the importance of the previous analysis focused on spatially targeted segment-scale interventions.

% =========================================================
% 4. DISCUSSION
% =========================================================
\section{Discussion}

% (note in original: 721 WORDS vs. 792 reference paper)

\subsection{Trip-length effects and distribution of cooling benefits}

Tree planting substantially reduces the total distance ridden under heat-stress conditions, yet it produces relatively smaller decreases in the share of trips exceeding critical UTCI thresholds ($\geq 32^{\circ}$C and $\geq 38^{\circ}$C). This suggests that the intervention benefits longer trips more than short, localized ones. Because longer trips often traverse high-use segments, many 
overlapping with the targeted, heat-exposed segments, these riders experience greater cumulative cooling.

Accordingly, the cooling benefits of tree planting appear concentrated along major, high-use segments, routes typically covered by longer trips, while shorter, local trips see more limited reductions in heat exposure. Even though these interventions are highly effective at reducing total heat-exposed kilometers ridden, additional measures are needed to bring all trips fully 
below the danger-zone thresholds. Complementary strategies such as cooling pavements could help address remaining hot spots and ensure that benefits extend to riders making short, intra-neighborhood trips. Together, these measures would provide a more comprehensive improvement in cyclist thermal comfort across the network.

\subsection{Efficiency and planning implications of targeted interventions}

Comparing hypothetical cooling scenarios shows that applying stronger cooling to a smaller number of street segments is more efficient than weaker, broader interventions. Cooling the 150 hottest segments by 10$^{\circ}$C reduces heat-exposed kilometers by 27.3\%, outperforming 400 segments at 5$^{\circ}$C (16.9\%) and 1{,}000 segments at 5$^{\circ}$C (24.9\%). This nonlinear response reveals diminishing returns once interventions extend beyond the most critical segments. From a planning perspective, concentrating resources on high-exposure areas can deliver greater overall reductions in rider heat burden at lower cost. In practice, prioritizing a limited number of ``cool corridor'' investments, rather than spreading efforts citywide, optimizes both cooling 
efficiency and resource allocation.

Compared with randomized tree-planting, targeted interventions also make more effective use of resources. While uniform canopy expansion cools the city overall, its impact is diffuse and requires substantial planting to achieve modest reductions. In contrast, focusing trees along the most heat-exposed, high-use segments achieves a comparable 19\% reduction in heat-exposed 
kilometers, roughly equivalent to an idealized 4$^{\circ}$C cooling, using far fewer resources. Spatially targeted greening thus provides a higher return on investment, maximizing thermal relief for riders while minimizing the scale and cost of intervention.

Aside from spatial efficiency, tree planting also provides a temporally optimized form of cooling. The diurnal analysis shows that added canopy produces the greatest reductions in UTCI during midday and early afternoon, the same hours when heat stress peaks. Although minor nocturnal warming occurs due to heat storage within canopy layers, its magnitude ($<0.5^{\circ}$C) and 
timing (at night, when ambient temperatures are lowest) are negligible compared with the daytime cooling observed between 09:00 and 16:00. This corresponds to an effective cooling equivalent of roughly 3--5$^{\circ}$C, depending on the hour, during the most critical period for human exposure. In other words, trees cool the city when it matters most: the hotter the conditions, the greater their cooling impact. This self-reinforcing response makes urban vegetation a realistic and time-sensitive mitigation mechanism for heat-stress adaptation.

Beyond direct cooling efficiency, greening these segments could also create a positive feedback loop for active mobility. Tree-lined ``cool corridors'' would not only reduce thermal stress but also improve aesthetic quality and comfort, likely attracting more cyclists and increasing overall ridership. This self-reinforcing dynamic, where more comfort leads to greater use, further justifying investment, underscores the broader co-benefits of targeted green 
infrastructure in promoting sustainable, heat-resilient urban transport.

\subsection{Socioeconomic patterns of rider heat exposure}

Another important consideration is the well-documented disparity in urban heat exposure across income levels. Low-income neighborhoods are known to experience disproportionate thermal burdens due to limited vegetation, higher impervious surface coverage, and less access to cooling infrastructure~\cite{ref2,ref12,ref13}. This pattern is also evident in our results (see Figure~7): trips through lower-income areas experience higher daytime UTCI values, particularly under direct solar radiation. Downtown districts appear cooler during the day due to building shadowing and greater vegetation, while peripheral, lower-income neighborhoods remain more exposed.

This shows that thermal stress is not only a function of where people live, but also of how and where they move through the city. Future work could explicitly integrate socioeconomic indicators into the prioritization framework developed here, ensuring that targeted mitigation not only maximizes overall cooling efficiency but also directs resources toward communities most affected by extreme heat.

% =========================================================
% 5. CONCLUSIONS
% =========================================================
\section{Conclusions}

By coupling high-resolution urban microclimate modeling with detailed mobility data, this study provides the first citywide quantification of heat exposure among micromobility users and evaluates how strategic tree planting can maximize cooling efficiency across the urban network. The results demonstrate that heat exposure among micromobility users is highly concentrated along a small subset of street segments that carry most of the rider traffic. 
Targeted cooling along these high-use, high-exposure routes yields the greatest benefits, reducing total heat-exposed kilometers by more than half in the most intensive scenarios.

Compared with randomized citywide greening, spatially prioritized tree planting delivers far higher cooling efficiency, achieving an effect equivalent to 3--5$^{\circ}$C artificial cooling with far fewer resources. Moreover, the diurnal analysis shows that trees provide maximum cooling during the hottest hours of the day, when both heat stress and riding activity peak, while only 
minimally warming the city at night.

Although a comprehensive equity assessment lies beyond the scope of this work, our analyses reveal a persistent negative correlation between income and rider heat exposure, suggesting that targeted greening efforts could also yield social co-benefits in the city’s hottest, most vulnerable areas.

Together, these findings highlight the importance of spatial and temporal prioritization in urban heat mitigation planning. As cities expand active mobility infrastructure in a warming climate, data-driven approaches such as this can be integrated with predictive mobility models \cite{steentoft2024,yang2025up} to enable the design of thermally resilient and equitable cycling networks.

% =========================================================
% REFERENCES (numbered, as in the Word version)
% You can either keep this thebibliography environment
% or move these into a .bib file and use BibTeX.
% =========================================================

% =========================================================
% ACKNOWLEDGEMENTS
% =========================================================

\section*{Acknowledgements}
The authors thank Marco Giometto and Emma Corbett for helpful discussions.

% =========================================================
% FIGURES (converted from the PDF pages)
% =========================================================

\begin{figure}[p]
    \centering
    \begin{tikzpicture}
        % ---- background map ----
        \node[anchor=south west, inner sep=0] (bg)
            at (0,0)
            {\includegraphics[width=0.95\textwidth]{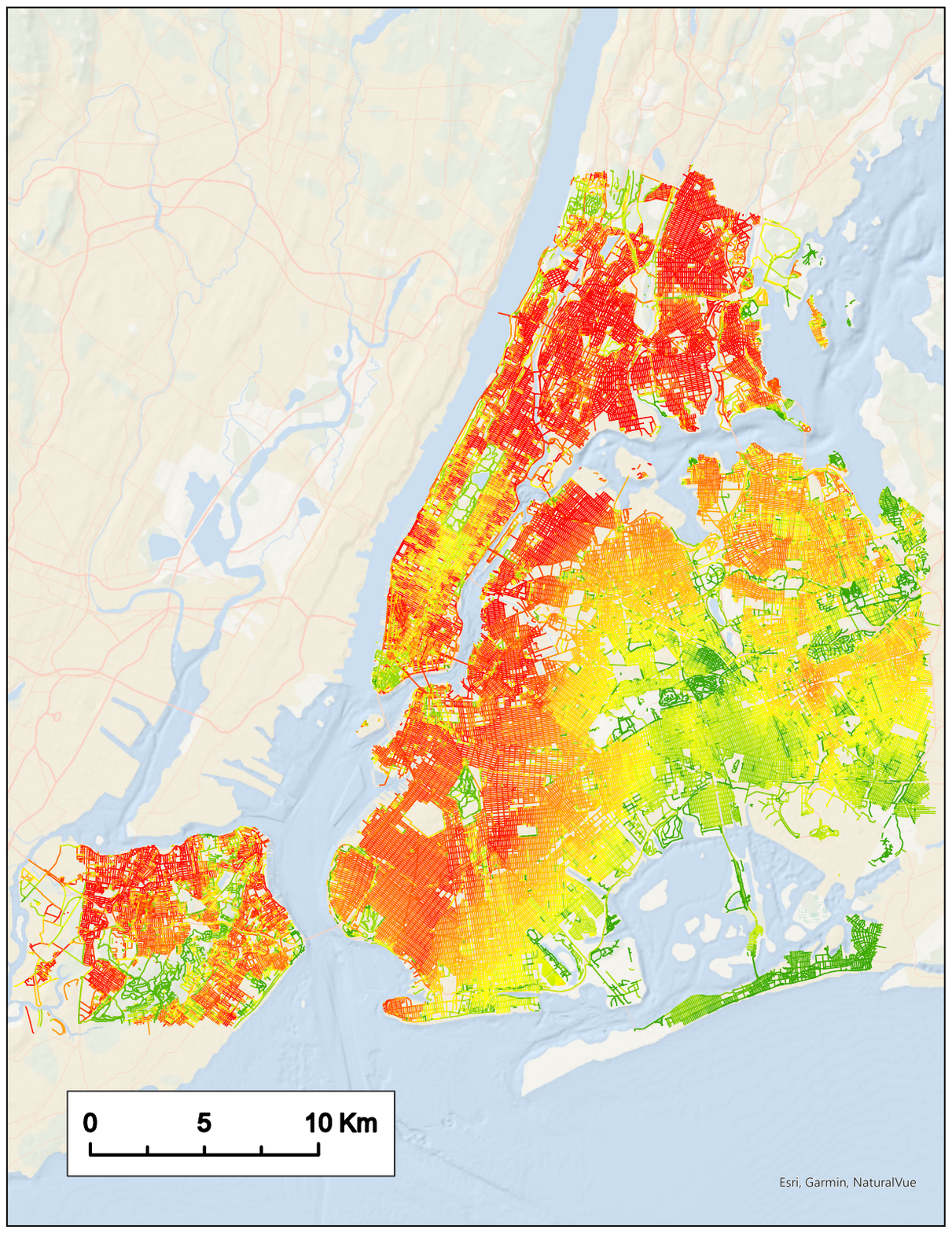}};

        % ---- inset map (top-left corner) ----
        % adjust xshift, yshift, and width
        \node[
            anchor=north west,
            inner sep=1pt,
            draw=black,          % draw border
            line width=0.375pt
        ] (inset)
            at ([xshift=0.024\textwidth,yshift=-0.029\textwidth]bg.north west)
            {\includegraphics[width=0.375\textwidth]{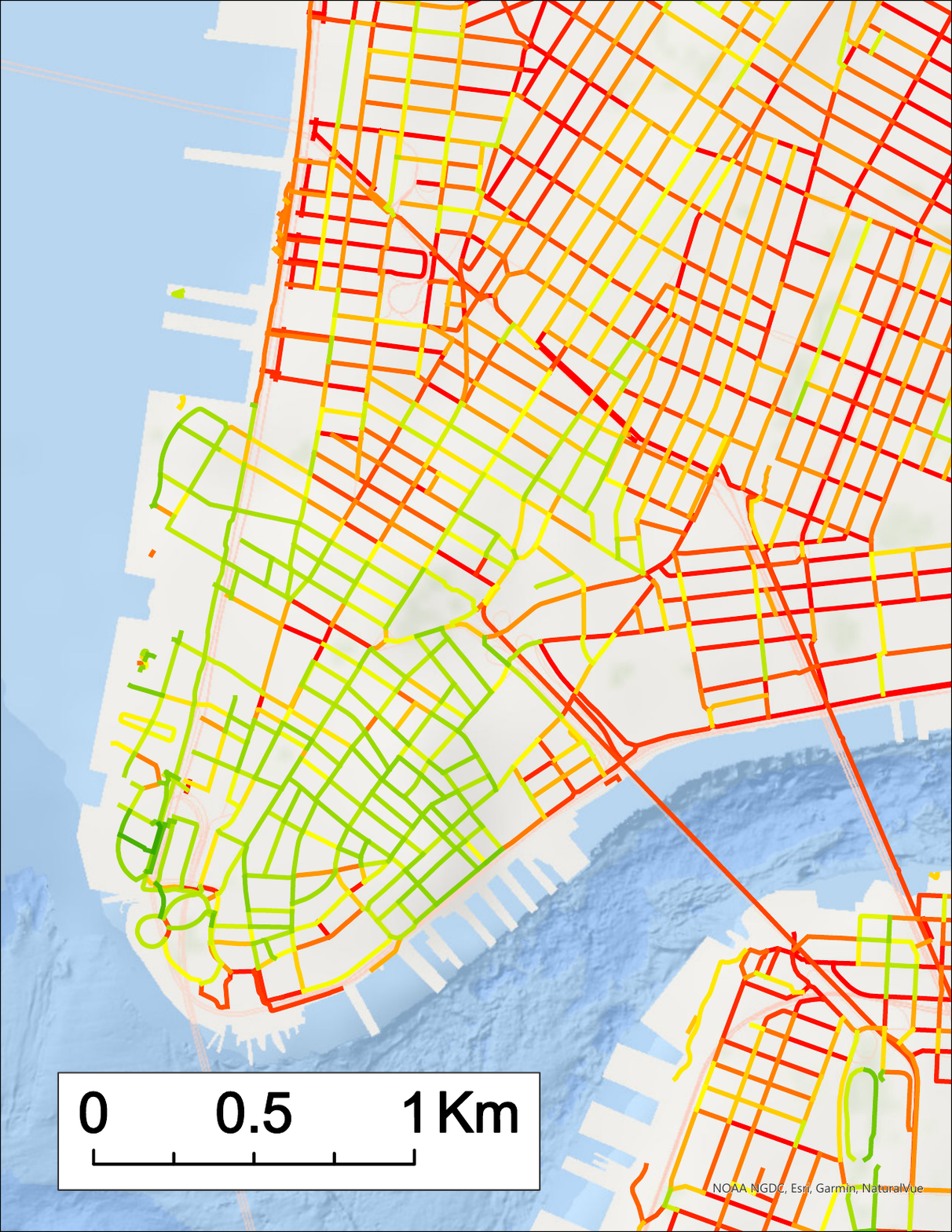}};
    \end{tikzpicture}

    \caption{Street-level map of UTCI values across New York City. The inset presents a detailed view of the street network in Lower Manhattan. The integration of WRF-BEP and SOLWEIG provides a high-resolution representation of microclimatic variability, capturing features such as the oceanic cooling effect along southern Queens and coastal Brooklyn, where sea-breeze circulation lowers UTCI values relative to inland neighborhoods.}
    \label{fig:1}
\end{figure}

\begin{figure}[p]
    \centering
    \includegraphics[width=0.95\textwidth]{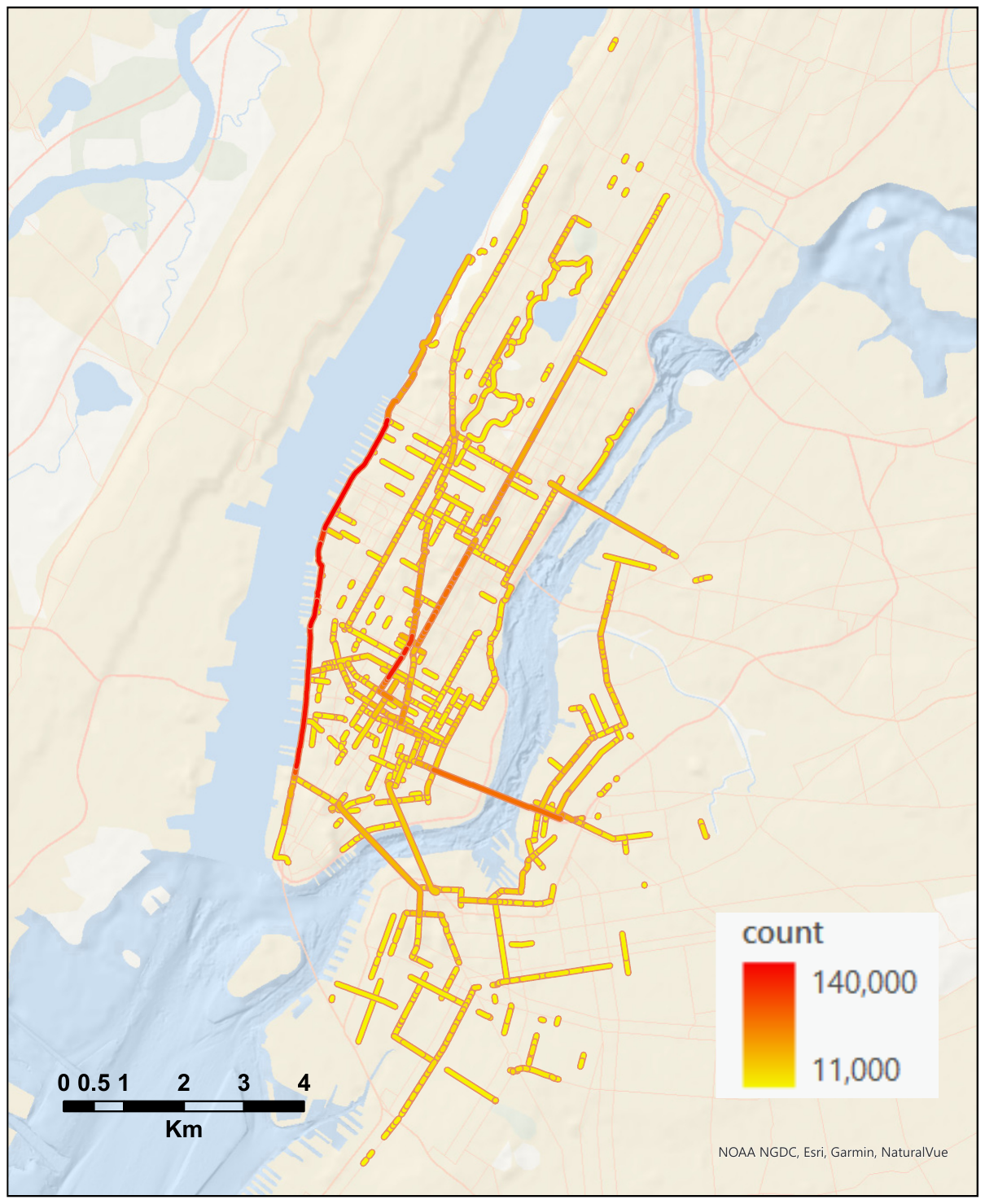}
    \caption{Global distribution of the top 1,000 heat-exposed street segments. For each hour of the day, segments with UTCI values exceeding 32$^{\circ}$C were identified. For these segments, the number of Citi Bike trips passing through them was summed across all hours to determine the most heat-exposed segments citywide. Warmer colors indicate higher trip counts across heat-stressed segments. Major corridors include the Hudson River Greenway, Park Avenue, the East River bridges, and sections of Broadway Avenue.}
    \label{fig:2}
\end{figure}

\begin{figure}[p]
    \centering
    % ---------- top row ----------
    \begin{minipage}{0.48\textwidth}
        \centering
        \includegraphics[width=\linewidth]{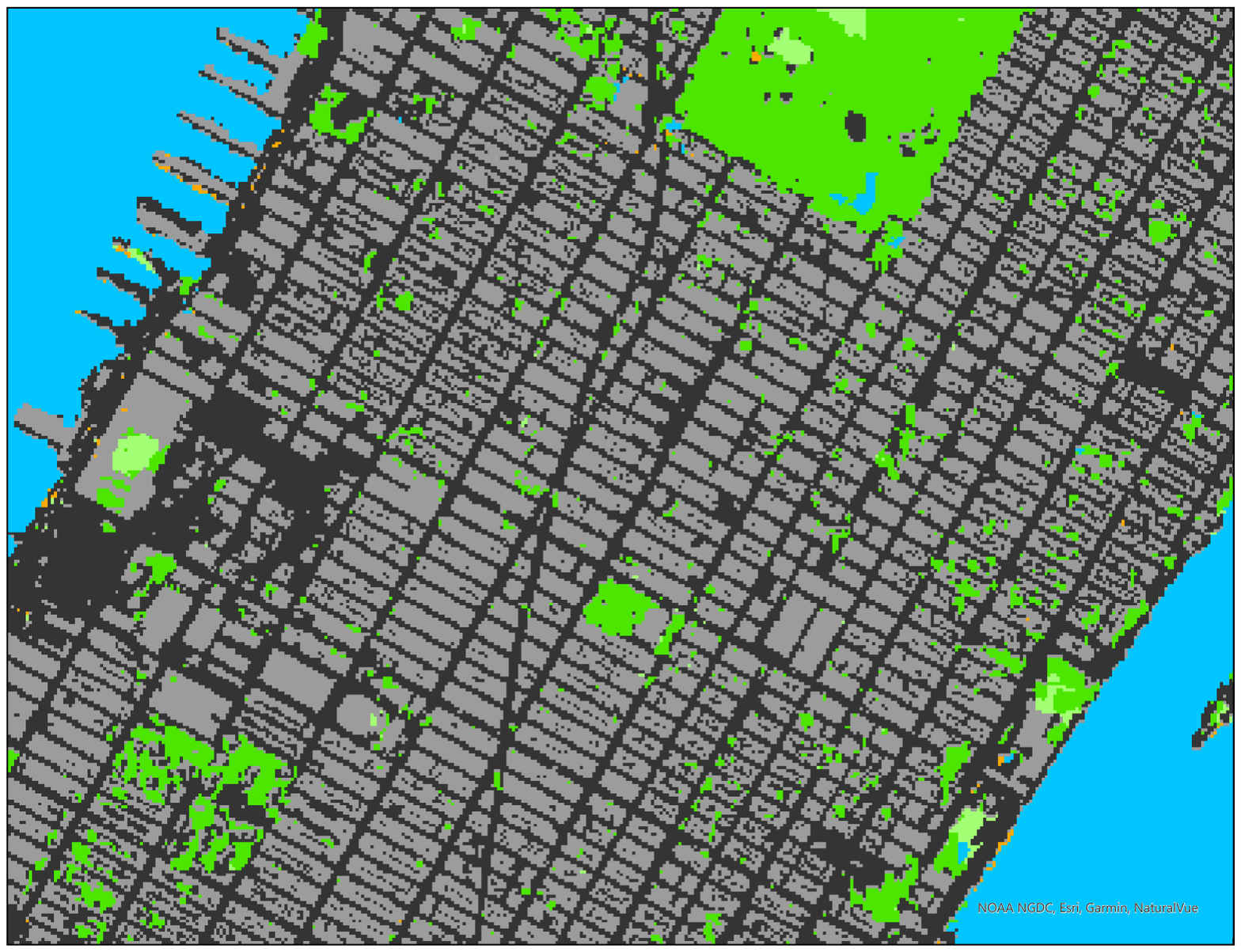}
        \\[-0.5em]
        {\small (a) Base scenario}
    \end{minipage}
    \hfill
    \begin{minipage}{0.48\textwidth}
        \centering
        \includegraphics[width=\linewidth]{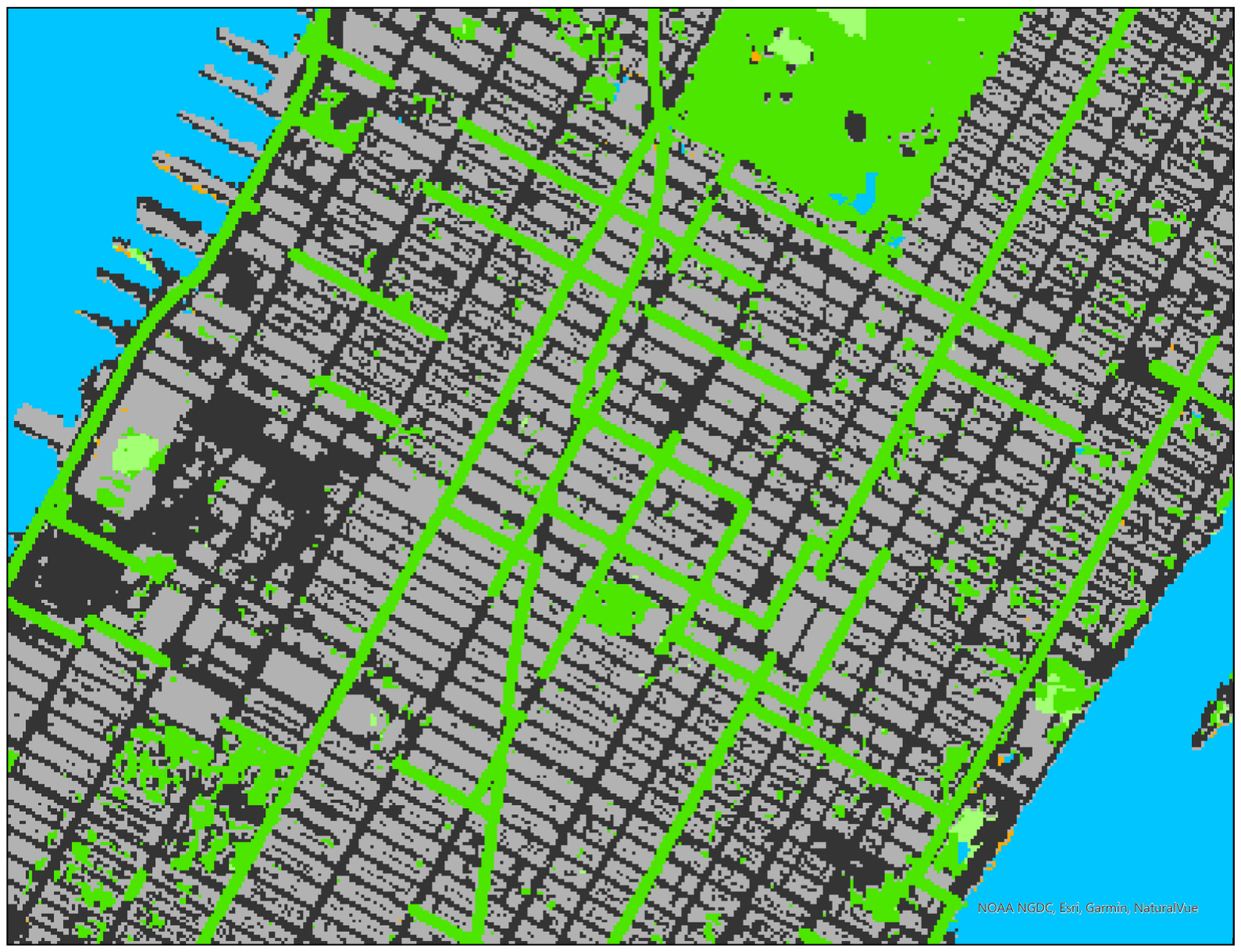}
        \\[-0.5em]
        {\small (b) Tree-planting scenario}
    \end{minipage}
    \vspace{1em}
    % ---------- bottom row ----------
    \begin{minipage}{0.9\textwidth}
        \centering
        \includegraphics[width=\linewidth]{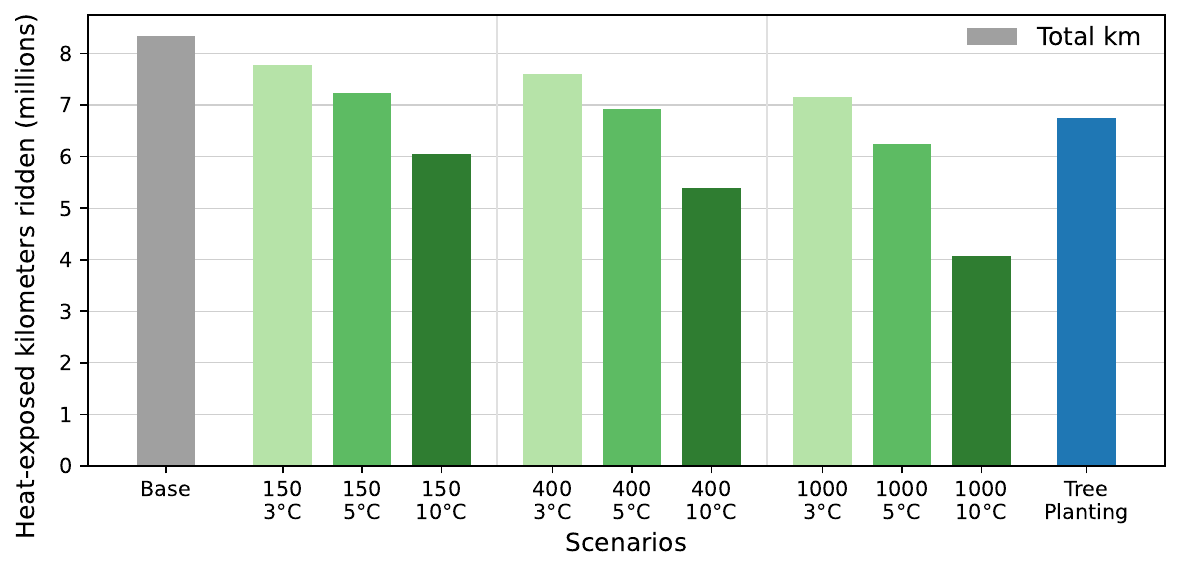}
        \\[-0.5em]
        {\small (c) Heat-exposed kilometers across scenarios}
    \end{minipage}

    \caption{Land-cover comparison for Midtown Manhattan under the (a) \textit{Base} and (b) \textit{Tree-Planting} scenarios, showing added canopy cover across the hottest 1,000 street segments. Panel (c) shows the resulting reduction in heat-exposed kilometers (UTCI $\geq 32^{\circ}$C) across all cooling and intervention scenarios. The gray bar represents baseline exposure, green bars represent theoretical cooling scenarios, and the blue bar represents the targeted tree-planting intervention.}
    \label{fig:3}
\end{figure}

\begin{figure}[p]
    \centering
    \includegraphics[width=\textwidth]{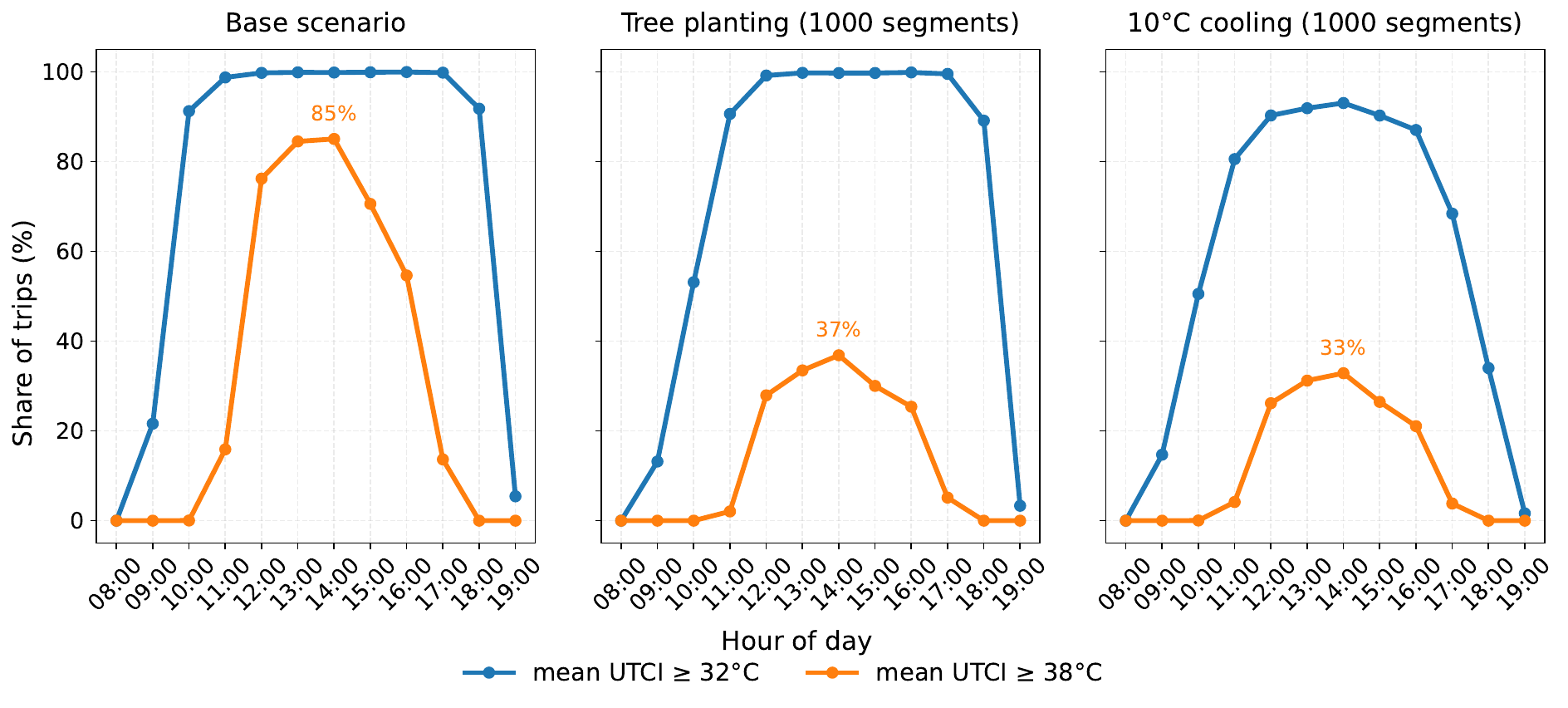}
    \caption{Danger-zone comparison. Hourly share of Citi Bike trips exceeding the UTCI danger-zone thresholds ($\geq 32^{\circ}$C and $\geq 38^{\circ}$C) for three modeled conditions: the base scenario (left), tree-planting scenario (middle), and 10$^{\circ}$C cooling of the top 1,000 heat-exposed segments (right).}
    \label{fig:4}
\end{figure}

\begin{figure}[p]
    \centering

    \includegraphics[width=0.9\textwidth]{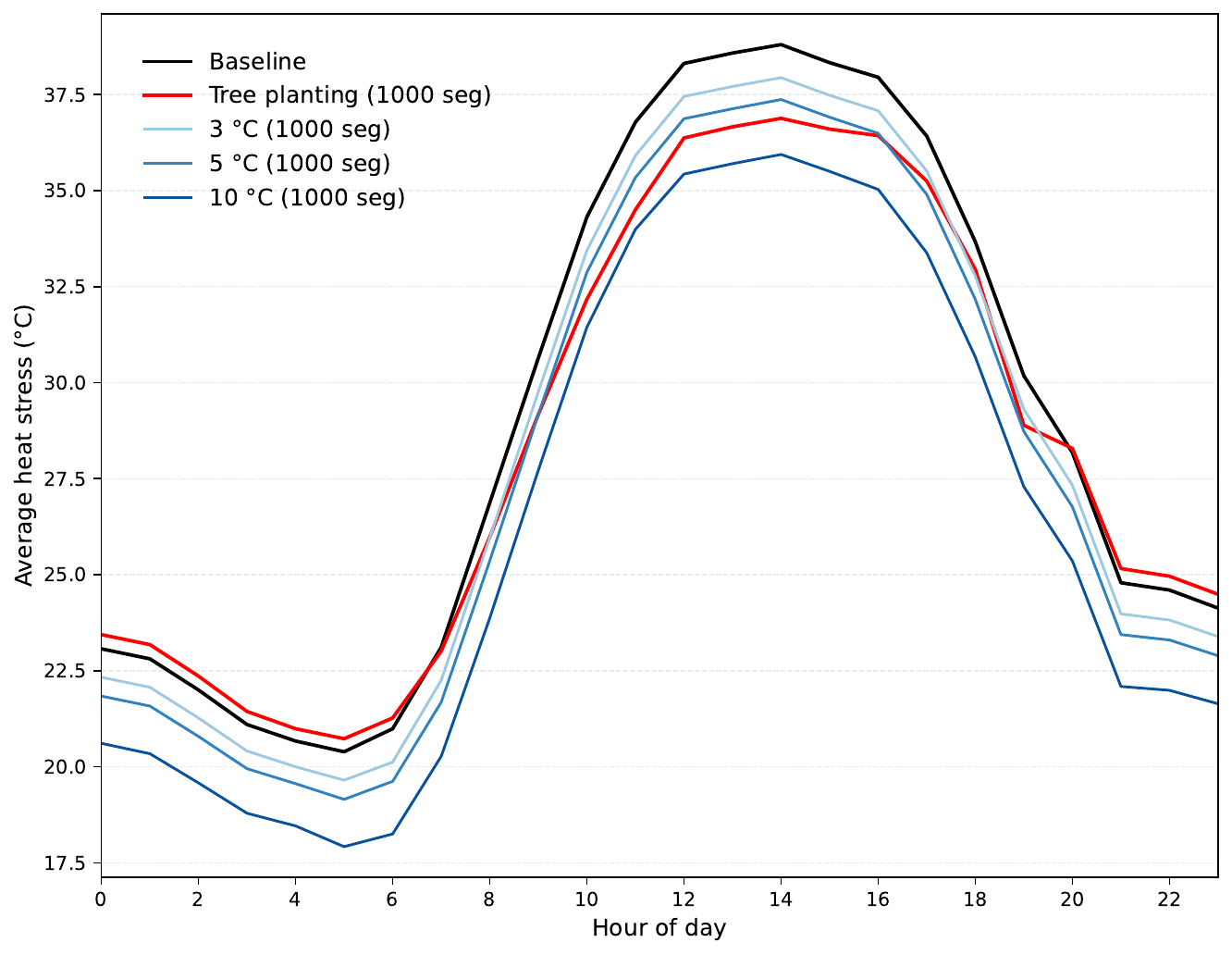}

    \includegraphics[width=0.9\textwidth]{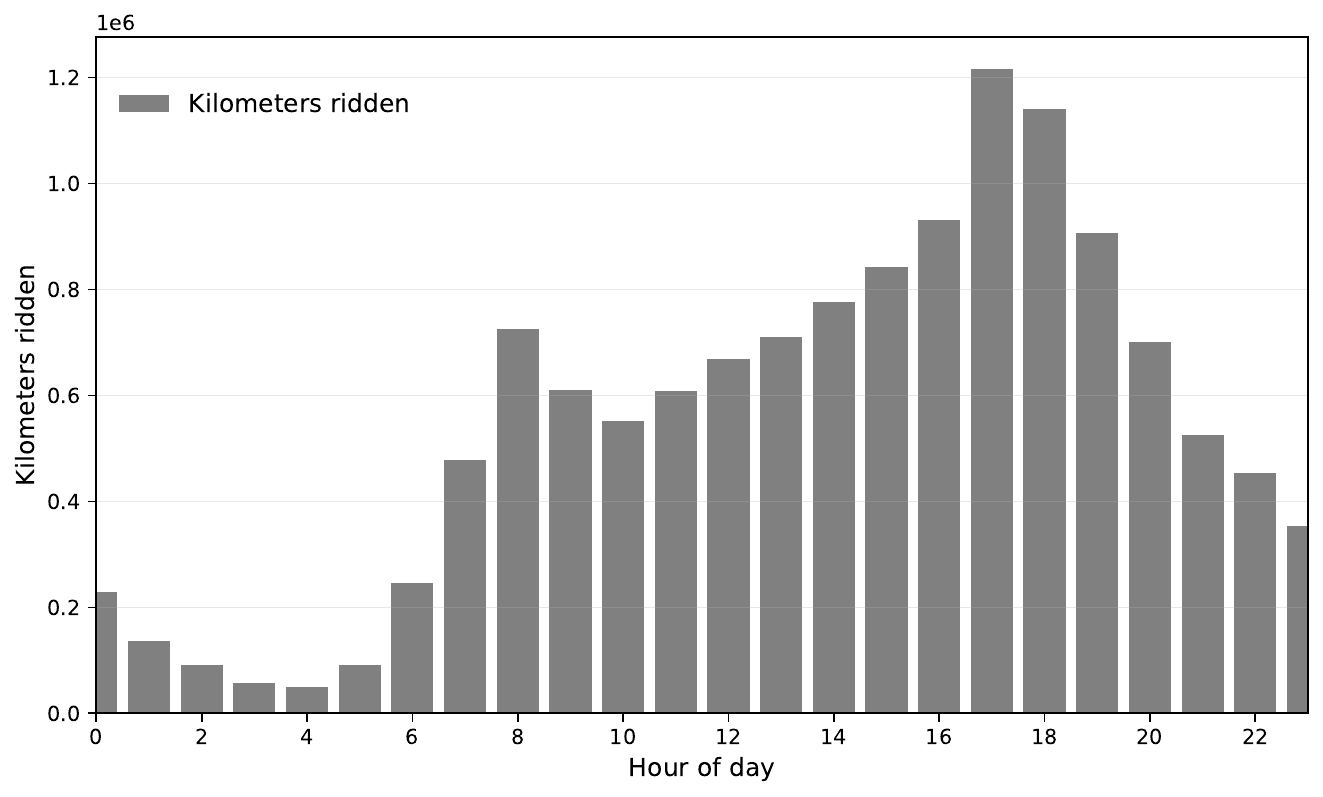}

    \caption{Combined hourly heat-exposure patterns and mobility demand for the top 1,000 tree-planting segments. \textit{(Top)} Hourly average heat stress (°C) under the baseline, tree-planting, and idealized-cooling scenarios. Tree planting produces modest nocturnal warming (+0.3--0.4°C) due to canopy heat storage but provides substantial daytime cooling of 1.5--2°C between 09:00 and 17:00, comparable to 3--5°C idealized-cooling equivalents. \textit{(Bottom)} Total kilometers ridden per hour.}
    \label{fig:5}
\end{figure}

\begin{figure}[p]
    \centering
    \includegraphics[width=\textwidth]{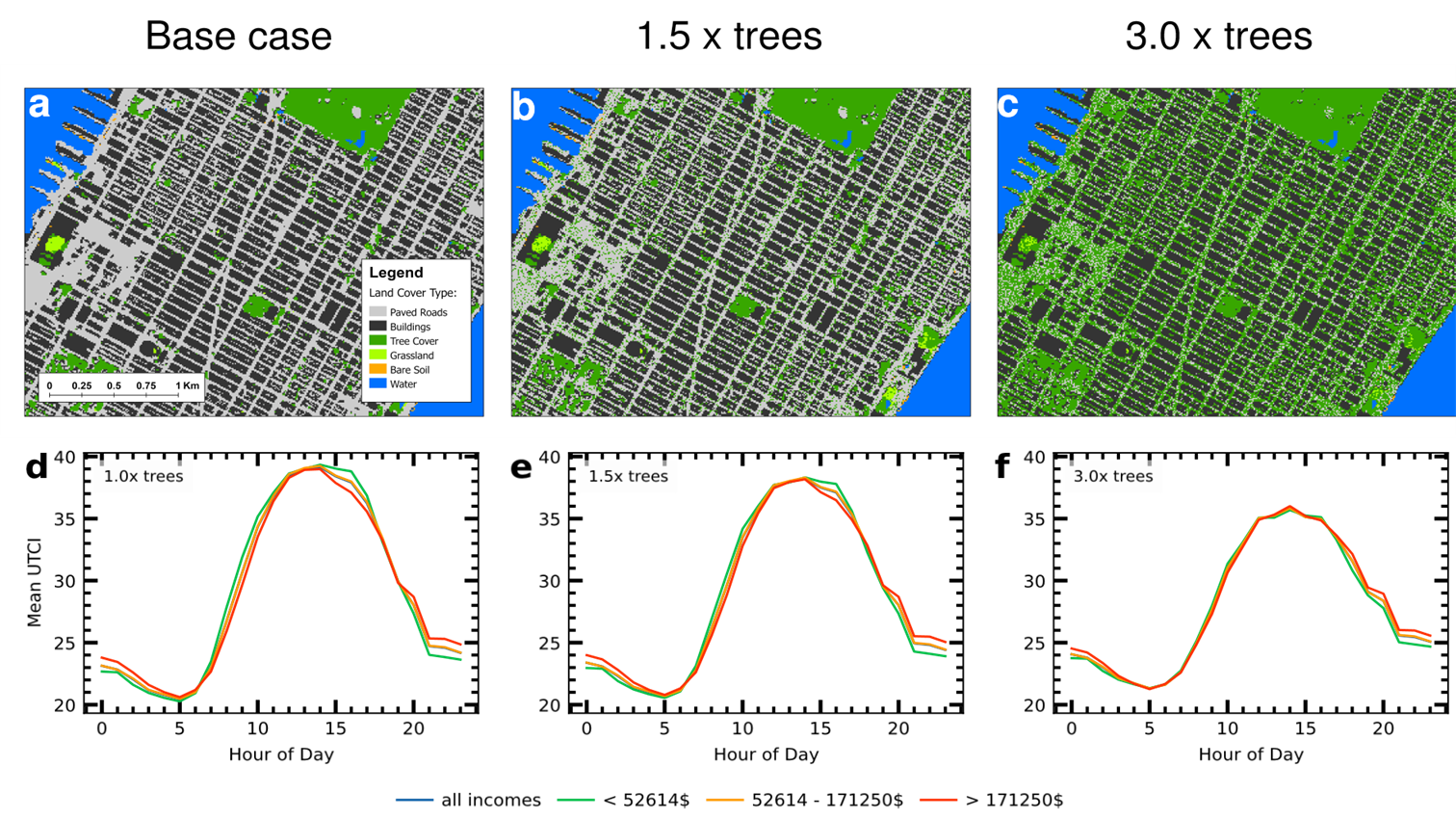}
    \caption{Citywide heat mitigation by random tree planting. (a–c) Land-cover maps illustrating increased tree density across Midtown Manhattan. The added trees are randomly distributed along streets and public spaces, comparing the current baseline (1.0×) with scenarios of 1.5× and 3.0× citywide canopy. (d–f) Corresponding diurnal profiles of mean UTCI across all NYC street segments under the same scenarios. As the canopy increases, the daytime UTCI decreases.}
    \label{fig:6}
\end{figure}

\begin{figure}[p]
    \centering

    % ---------- Top row: full-width map ----------
    \includegraphics[width=0.5\textwidth]{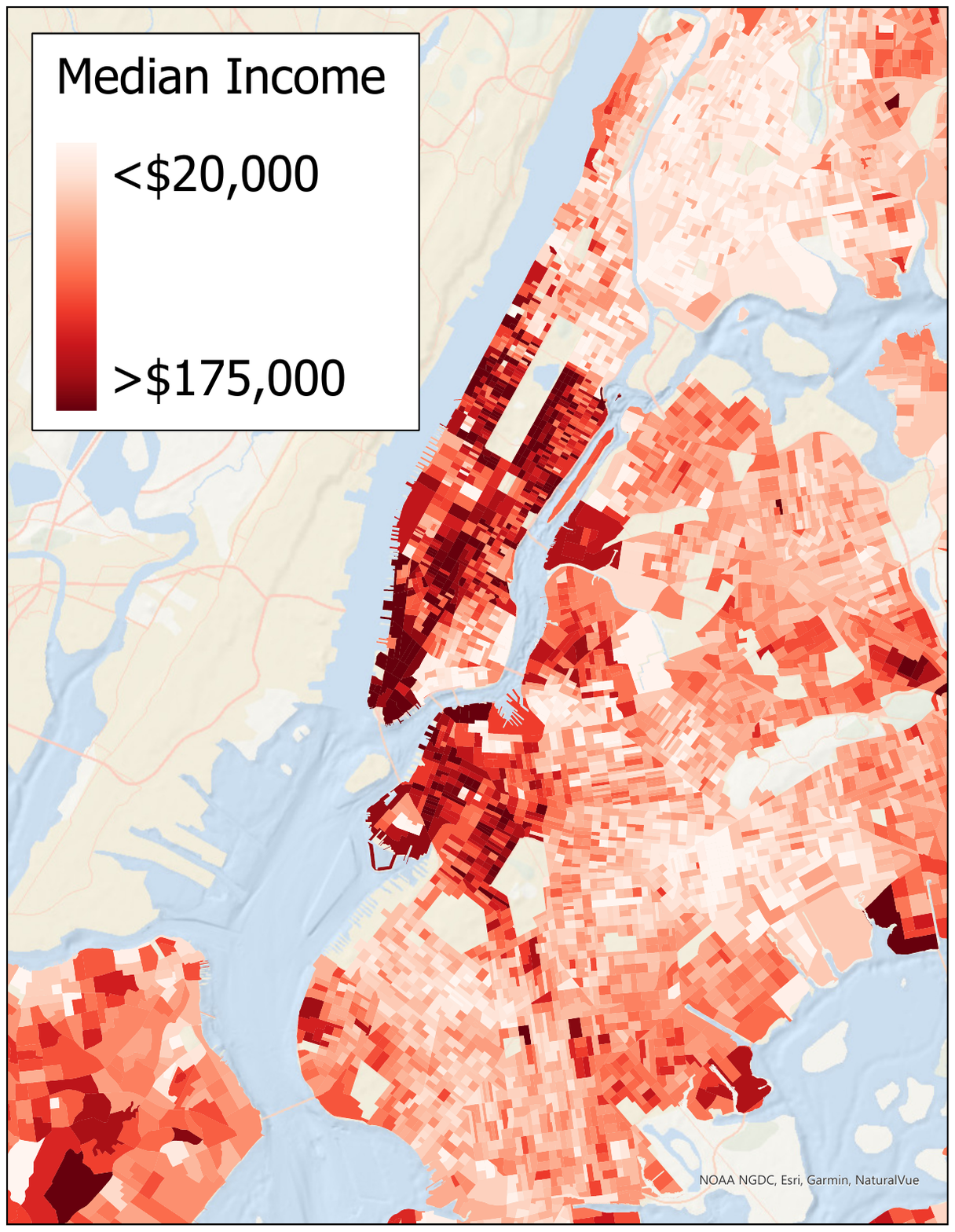}

    % ---------- Bottom row: two equal panels ----------
    \begin{minipage}{0.495\textwidth}
        \centering
        \includegraphics[width=\linewidth]{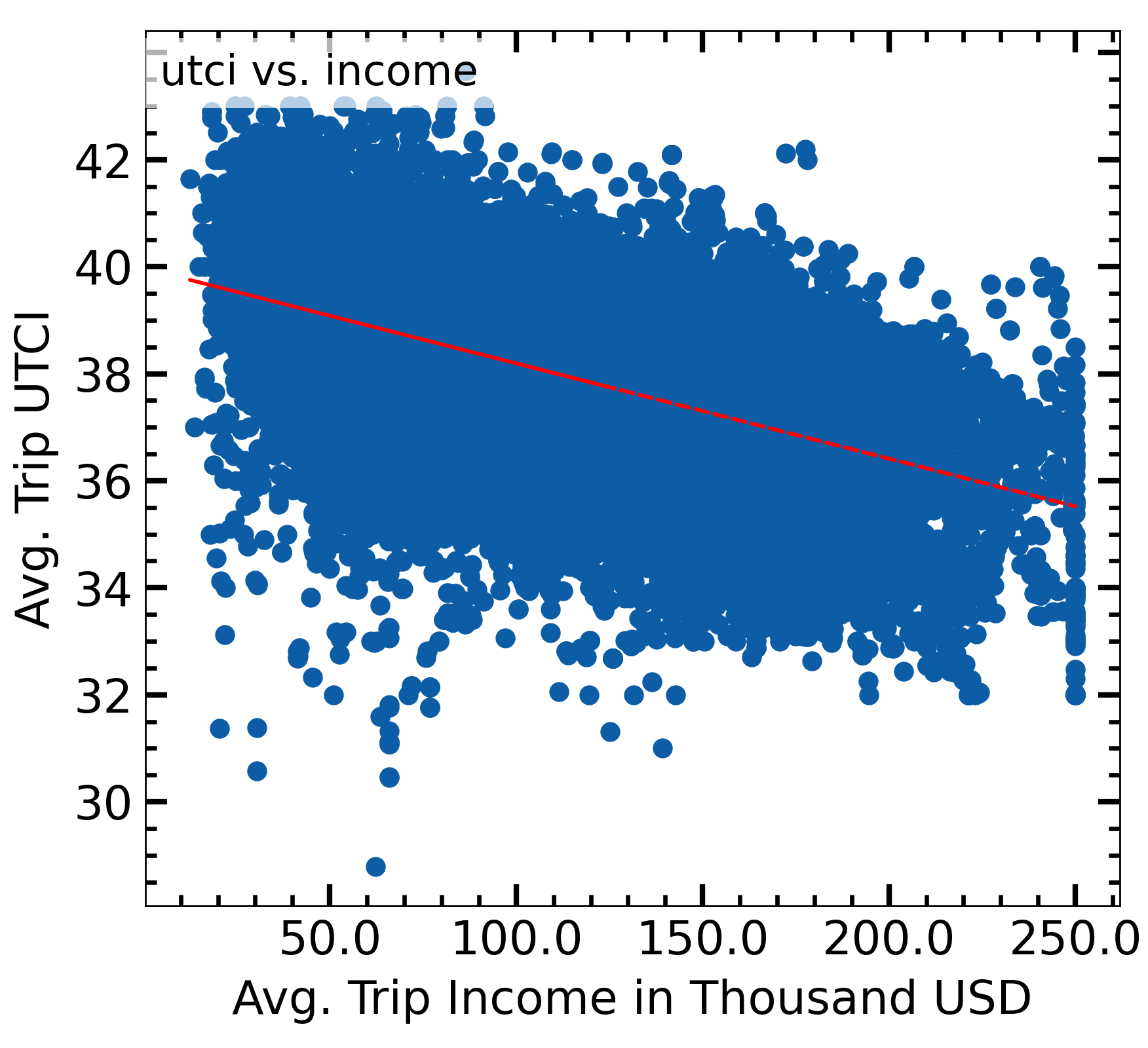}
    \end{minipage}
    \begin{minipage}{0.495\textwidth}
        \centering
        \includegraphics[width=\linewidth]{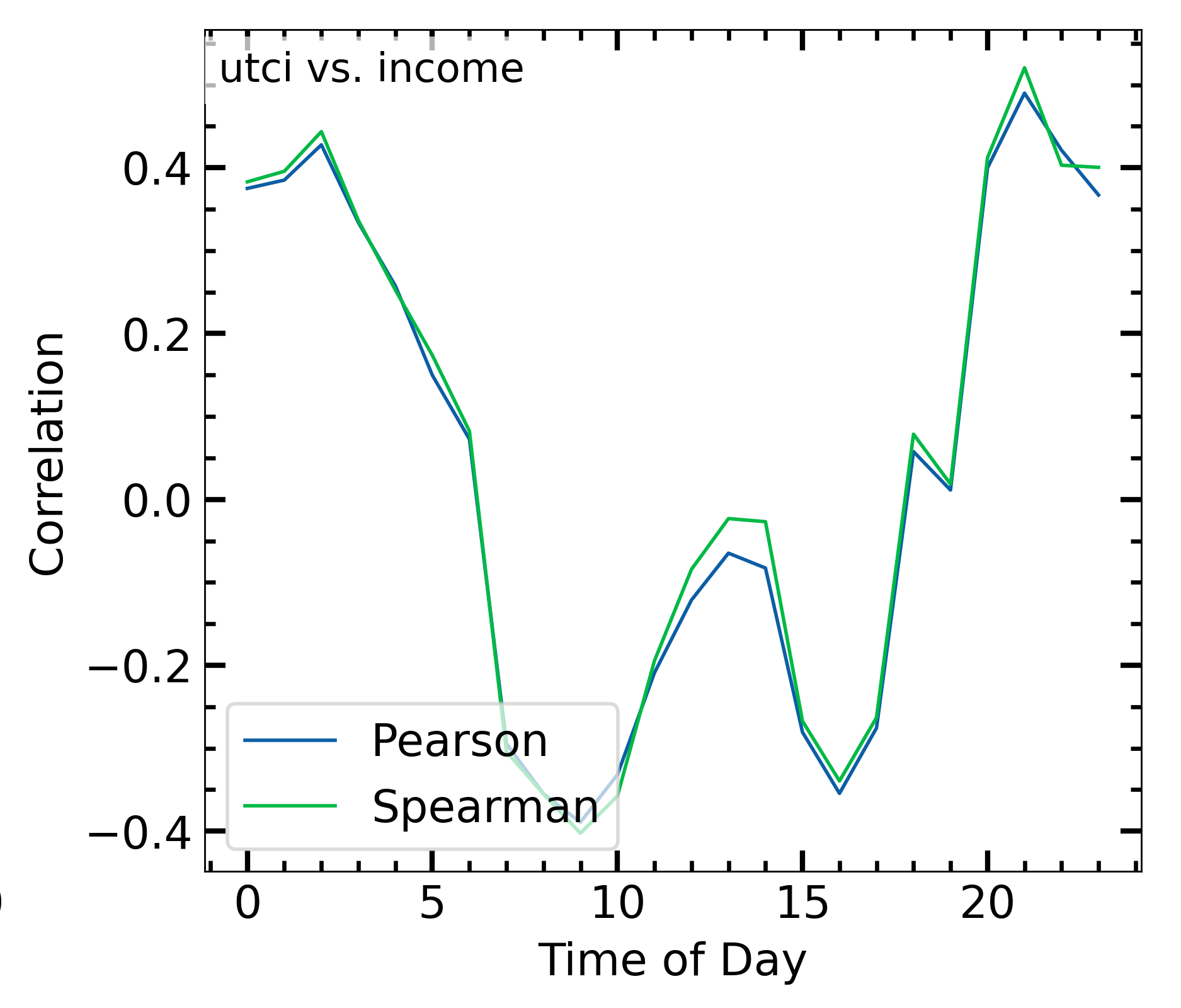}
    \end{minipage}

    \caption{Quantifying trip-based heat and income exposures. \textit{(top)} Census median household income mapped to each street segment. \textit{(bottom left)} Relationship between the Universal Thermal Climate Index (UTCI) and income for all road segments at noon. \textit{(bottom right)} Hourly correlation coefficients between income and UTCI across all street segments. Around noon, the correlation approaches zero because the high sun angle minimizes building shading, temporarily reducing the cooling advantage of high-income areas. For details on the calculation of average trip income, see Supplementary Information.}
    \label{fig:7}
\end{figure}

\end{document}